\documentclass[]{article}

\usepackage{graphicx}

\usepackage{texdraw}
\usepackage{epic,eepic}
\usepackage[english]{babel}
\usepackage[centertags]{amsmath}
\usepackage{amsfonts}
\usepackage{amssymb}
\usepackage{color}

\newtheorem{prop}{Proposition}

\newcommand{\ba}{\begin{array}}
\newcommand{\ea}{\end{array}}

\newcommand{\be}{\begin{equation}}
\newcommand{\ee}{\end{equation}}

\newcommand{\dis}{\displaystyle}

\newcommand{\x}{w}

\def\mref#1{(\ref{#1})}
\def\eqref#1{(\ref{#1})}

\begin{document}
\title{On non-multiaffine consistent-around-the-cube lattice equations}
\author{
Pavlos Kassotakis
\thanks{\emph{Present address:} Department of Mathematics and Statistics University of Cyprus,
P.O Box: 20537, 1678 Nicosia, Cyprus;
\newline \emph{e-mails:} {\tt kassotakis.pavlos@ucy.ac.cy, pavlos1978@gmail.com}}
\and
Maciej Nieszporski
\thanks{\emph{Present address:} Katedra Metod Matematycznych Fizyki, Uniwersytet Warszawski,
ul. Ho\.za 74, 00-682 Warszawa, Poland;
\emph{e-mail:} {\tt maciejun@fuw.edu.pl}}
}
\maketitle

\begin{abstract}
We show that integrable involutive maps, due to the fact they admit three integrals in separated form, can give rise
to equations,
%{\color{red} (or in general  to  correspondences)},
 which are consistent  around the cube and which
are not in the multiaffine form  assumed in papers \cite{ABS,ABS2}.
%{\color{red} (we propose how to extend the notion of consistency  around the cube to  correspondences)}.
Lattice models, which are discussed  here,
%{\color{red} (correspondences)}
are related to the lattice potential KdV equation by nonlocal transformations (discrete quadratures).
\end{abstract}

{\bf Pacs:} 02.30.Ik, 05.50.+q

{\bf Keywords:} Integrable lattice equations, Yang-Baxter maps, Consistency around the cube

{\bf Mathematics Subject Classification:} 82B20, 37K35, 39A05

%%%%%%%%%%%%%%%%%%%%%%
\section{Introduction}
%%%%%%%%%%%%%%%%%%%%%%
For a long time the electromagnetic potentials, $\Psi$ and {\bf A} were considered as convenient purely mathematical concepts
without physical significance, so to say phantoms, and
only the electromagnetic fields, {\bf E} and {\bf B}, were considered to be physical.
It was thought that a change of potential that did not affect the fields did not  cause measurable effects.
The discovery of the Aharonov-Bohm effect changed the situation.
The investigation of a system of equations together with its potential image is nowadays  more than merely searching for a convenient way of description.

Here we concentrate on the fact that one of the most important nonlinear difference equations, the lattice potential Korteweg-de Vries (lpKdV) equation
\mref{xi} (see \cite{NQC,NC}), arises as a potential version
of a more fundamental system\footnote{The word ``potential'' in the name of the equation lpKdV is some remnant from classical terminology used in the theory of integrable discretizations of integrable systems and should not be  confused with the meaning of
the scalar ``potential'' of a discrete vector field which
we use throughout the paper.}
{\small
\begin{equation}
\label{uv}
u(m+\frac{1}{2}, \! n+1)
 \! = \! \frac{p(m+\frac{1}{2})-q(n+\frac{1}{2})}{u(m \! + \! \frac{1}{2}, \! n) \! - \! v(m, \! n \! + \! \frac{1}{2})}-v(m, \! n+\frac{1}{2}),
\quad v(m+1, \! n+\frac{1}{2}) \! = \!  \frac{p(m+\frac{1}{2})-q(n+\frac{1}{2})}{u(m \! + \! \frac{1}{2}, \! n) \! - \! v(m, \! n \! + \! \frac{1}{2})}-u(m+\frac{1}{2}, \! n),
\end{equation}}
where $(m,n)\in {\mathbb Z}^2$, so the function $u$ can be regarded as given on horizontal edges of ${\mathbb Z}^2,$
  $v$  on vertical edges and $p$, $q$ are given functions of a single variable (see Section \ref{main}).

 The system \mref{uv}  originates  from papers  where the interrelation between discrete integrable systems and set theoretical solutions of the Yang-Baxter equation (the so called Yang-Baxter maps) were
investigated (see Section \ref{yb} for details and a list of references). A surprise and the main result of this paper is that the fundamental system
 \mref{uv} has  additional  potentials and moreover the system
can be easily extended to higher dimensions \mref{ui}.

To be more specific the  system  \mref{uv} allows us to introduce  three potentials defined on vertices of the lattice, say $x$, $z$ and~$f$,
defined respectively by
\begin{equation}
\label{1}
\begin{array}{l}
u(m+\frac{1}{2},n)= x(m+1,n)-x(m,n) , \quad   v(m,n+\frac{1}{2}) =  x(m,n+1)-x(m,n)
\end{array}
\end{equation}
\begin{equation}
\label{2}
\begin{array}{l}
\!\!\! \!\!\! \!\!\! u(m+\frac{1}{2},n)^2 = z(m+1,n)+z(m,n)+p(m+\frac{1}{2}) , \quad   v(m,n+\frac{1}{2})^2 =  z(m,n+1)+z(m,n)+q(n+\frac{1}{2})
\end{array}
\end{equation}
{\small
\begin{equation}
\label{3}
\begin{array}{l}
\!\!\! \!\!\! \!\!\!  \!  \!\!\! \!\!\!\!\!\!u(m \! + \! \frac{1}{2}, \! n) ^3 \! - \! 3p(m \! + \! \frac{1}{2}) u(m \! + \! \frac{1}{2}, \! n) \! = \! f(m \! + \! 1, \! n) \! - \! f(m, \! n), \quad
 v(m, \! n \! + \! \frac{1}{2}) ^3 \! - \! 3q(n \! + \! \frac{1}{2}) v(m, \! n \! + \! \frac{1}{2}) \!  =  \! f(m, \! n \! + \! 1) \! - \! f(m, \! n)
\end{array}
\end{equation}}
%{\color{red} where   \[a(m+\frac{1}{2})-b(n+\frac{1}{2})=3[q(n+\frac{1}{2})-p(m+\frac{1}{2})].\]}

Having in mind the philosophical repercussions that the discovery of Aharonov-Bohm effect triggered,
we believe that this nontrivial arbitrariness in introducing the scalar potential will be interesting in itself.
However, we focus here on relations that the potentials satisfy
and also we aim to contribute to the modern area of
discrete integrable systems related to the notion of the consistency around the cube property \cite{ABS,FN,FN-KN}.
Namely, in the case of the potential $f$ by confining ourselves to real-valued functions and  everywhere negative parameters $p$ and $q$, we get an equation on the lattice
\begin{equation}
\label{realcub}
 f_{12} =   f+  {\dis (p-q)\left[v-u+\frac{f_1-f_2}{(u-v)^2} +\frac{(p-q)^2}{(u-v)^3}\right]}
 \end{equation}
 $$
 \begin{array}{ll}
{\scriptsize u:= \sqrt[3]{\frac{f_1  - f}{2}+\sqrt{ \frac{(f_1  -  f)^2}{4}- p^3} }+\sqrt[3]{\frac{f_1 -  f}{2}-\sqrt{ \frac{(f_1  -  f)^2}{4}- p^3} }, } &
{\scriptsize v:= \sqrt[3]{\frac{f_2  -  f}{2}+\sqrt{ \frac{(f_2  -  f)^2}{4}- q^3} }+\sqrt[3]{\frac{f_2  -  f}{2}-\sqrt{ \frac{(f_2  -  f)^2}{4}- q^3} }, }
\end{array}
$$
where by subscript we denote the forward shift operator and at the same time we omit the independent variables in formulas e.g. $u(m+\frac{1}{2},n+1):=u_2$, $v(m+1,n+\frac{1}{2})=v_1$, $f(m+1,n+1)=f_{12}$ (see Figure \ref{fig:building-latt}).
To the best of our knowledge  equation \mref{realcub} is the first example of a 2D non-multiaffine consistent-around-the-cube lattice equation.
 We discuss also here a more complicated case of complex-valued functions.

We start the paper  with the basic  relations that the potentials satisfy (Section \ref{lat-pot}).
Next we recall  some basic properties of the lpKdV equation  including its consistency-around-the-cube  property (Section \ref{kdv}),
followed by recalling some Yang-Baxter maps related to lpKdV (Section \ref{yb}). We stress, anticipating facts, that the Yang-Baxter property is not essential here.
The origin of the potentials, namely the existence of specific integrals  of the maps, is explained in  Section \ref{yb} as well.
The main results of the paper are in
Section \ref{main}
and our intention, while writing this letter, was to keep Section  \ref{main} self-contained.

\section{Lattice models associated with scalar potentials}
%%%%%%%%%%%%%%%%%%%%%%%%%%%%%%%%%%%%%%%%%%%%%%%%%%%%%%%%%%%%%%%%%%%%%%%%%%%
\label{lat-pot}

Here we determine the relations that the potentials $x$, $z$ and $f$ satisfy and we discuss in what sense these relations are integrable.
%These relations, as we shall see, are either equations or correspondences.
Specifically from \mref{uv}, \mref{1},  \mref{2} and \mref{3} we infer that the potentials obey
%\begin{itemize}
%\item %lattice potential KdV equation denoted in \cite{ABS} by $H1$
%\begin{equation}
%\label{xi}
%\begin{array}{l}
%(H1): \qquad \qquad \qquad \qquad
%[x(m+1,n+1)-x(m,n)][x(m+1,n)-x(m,n+1)]=p(m)-q(n),
%(x_{12}-x)(x_1-x_2)=p-q,
%\end{array}
%\end{equation}
%\item
%\begin{equation}
%\label{zi}
%\begin{array}{l}
%z(m+1,n+1)=z(m,n)+{\dis [p(m)-q(n)] \frac{z(m,n+1)-z(m+1,n)}{[u(m,n)-v(m,n)]^2}},
%z_{12}=z+{\dis (p-q) \frac{z_2-z_1}{(u-v)^2}},
%\end{array}
%\end{equation}
%\item
%\begin{equation}
%\label{fi}
%\begin{array}{l}
%\!\!\!\!\!\!\!\!\!\!\!\!\!\!\!\!\!\!\!\!\!\!\!\!\!\!\!\!\!\!\!\!\!\!\!\!\!\!\!\!\!\!\!\!\!\!\!\!\!
%f(m+1,n+1)=f(m,n)+[p(m)-q(n)]\left[v(m,n)-u(m,n)+\frac{f(m+1,n)-f(m,n+1)}{[u(m,n)-v(m,n)]^2} +\frac{[p(m)-q(n)]^2}{[u(m,n)-v(m,n)]^3}\right],
%f_{12}=f+(p-q)\left(v-u+\frac{f_1-f_2}{(u-v)^2} +\frac{(p-q)^2}{(u-v)^3}\right),
%\end{array}
%\end{equation}
%\end{itemize}

\begin{eqnarray}
\label{xi}
 (x_{12}-x)(x_1-x_2)=p-q,\\
\label{zi}
 z_{12}=z+{\dis (p-q) \frac{z_2-z_1}{(u-v)^2}},\\
\label{fi}
 f_{12}=f+(p-q)\left(v-u+\frac{f_1-f_2}{(u-v)^2} +\frac{(p-q)^2}{(u-v)^3}\right),
\end{eqnarray}

Equation \mref{xi}  is referred to as the lattice potential KdV equation \cite{NQC,NC} and is denoted in \cite{ABS} by $H1$.
We can consider the system consisting of equations
\mref{2}, \mref{zi} and  the one that consists of equations \mref{3} and \mref{fi} as relations imposed on functions $z$ and $f$ respectively.
Equations \mref{2} and \mref{3} viewed as the definition of the functions $u$ and $v$ in terms of $z$ and $f$ respectively are not single-valued in the case of consideration of complex-valued functions
(the single-valued real case of \mref{3} and \mref{fi} leads to equation \mref{realcub}).
It means that relations \mref{2}, \mref{zi} and  \mref{3}, \mref{fi}
are ``ill-defined'' and should be regarded rather as correspondences  than equations (we are grateful to the referees  for this comment).
Instead of that we suggest a different point of view. We consider both systems \mref{2}, \mref{zi} and  \mref{3}, \mref{fi} together with equation
\mref{uv}.

This turns  correspondences \mref{2}, \mref{zi} and  \mref{3}, \mref{fi}  into well-defined lattice models \mref{uv}, \mref{2}, \mref{zi} (we shall denote the model by G1, the model can be found in disguise in \cite{ves3}) and \mref{uv}, \mref{3}, \mref{fi} (we shall refer to this  model as F1) defined on both vertices and edges of a 2D square lattice (nD lattice in the multidimensional extension).  Similar models have also been  introduced in the recent work of Hietarinta and Viallet \cite{hi-via}.
 We postpone to Section \ref{main} a remark on proper initial value problem of models F1 and G1.

%It makes the system \mref{2}, \mref{zi} and  \mref{3}, \mref{fi} well defined with only exception

The question arises: in what sense are these systems  integrable?
It turns out that both G1 and F1 are related to the lpKdV (H1) equation though the dependence is not local. Namely, if
$x$ obeys the lpKdV then the function $z$ given by discrete quadratures
\begin{equation}
\label{nabz}
\begin{array}{l}
z(m+1,n)+z(m,n)=[x(m+1,n)-x(m,n)]^2-p(m), \\ z(m,n+1)+z(m,n)=[x(m,n+1)-x(m,n)]^2-q(n)
\end{array}
\end{equation}
satisfies  G1,
whereas the function $f$ given by
\begin{equation}
\label{nabf}
\begin{array}{l}
f(m+1,n)-f(m,n)=(x(m+1,n)-x(m,n))^3-3p(m+\frac{1}{2})(x(m+1,n)-x(m,n)), \\ f(m,n+1)-f(m,n)=(x(m,n+1)-x(m,n))^3-3q(n+\frac{1}{2})(x(m,n+1)-x(m,n))
\end{array}\end{equation}
is a solution of  F1. So we understand the integrability  of G1, F1 as the possibility of finding their solutions
from known solutions of an integrable system (lpKdV in this case)
by solving {\em linear} equations \mref{nabz}, \mref{nabf} respectively.
 We call the equations \mref{nabz} and  \mref{nabf} non-auto-B\"acklund transformations between \mref{xi} and G1, F1 respectively. However, there is a tendency to refer to such transformations as to Miura-type transformations (compare \cite{Decio,RG2-Sol}).
Note that non-auto-B\"acklund transformations between equations in the Adler-Bobenko-Suris (ABS) list  \cite{ABS} are presented in
\cite{James,Atk-pre,Decio}.
%\cite{James,Decio}.

%{\color{blue}Jameses papers  Note that according to  results of paper \cite{James,Atk-pre,Decio}
%there exists non-auto-B\"acklund transformations (or Miura transformations)
%between all equations of the set (H1)--(H3) and (Q1)--(Q3) of paper \cite{ABS}.}

Moreover,
as we already mentioned, equation H1  and  the  models G1 and F1 which are given on a two-dimensional lattice, can be extended to %system of compatible  equations
 a compatible system
on an n-dimensional lattice, see Section \ref{main}.
This property (of ``compatible extendibility'' to higher dimensions) as it has been pointed out in  \cite{DS,CDS,FN,FN-KN,ABS} is a hallmark of integrability.
In other words, following the terminology introduced in \cite{ABS},  equation H1 and   models G1 and F1 possess  the  property of consistency around the cube.

To the end of this section we mention that the three cases above can be combined giving rise to a family of potentials $\psi$
\begin{equation}
%\label{}
\begin{array}{l}
c_0+c_1 u +c_2 (-1)^{m+n} ( u^2 - p ) + c_3(u^3 + p u )= \psi_1-\psi, \cr
c_0+c_1 v +c_2 (-1)^{m+n} ( v^2 - q ) + c_3(v^3 + q v )= \psi_2-\psi,
\end{array}
\end{equation}
where  $c_i\in \mathbb{C},$ $i=0,\ldots 3.$ 

%%%%%%%%%%%%%%%%%%%%%%%%%%%%%%%%%%%%%%%%
\section{Consistency around the cube of the lattice potential KdV equation}
%%%%%%%%%%%%%%%%%%%%%%%%%%%%%%%%%%%%%%%%
\label{kdv}
Here we recall some essential facts concerning the consistency around the cube of the lattice potential KdV equation and explain how the models F1 and G1 differ from equations in the ABS list.
All the formulas can be found in the classical paper by
Wahlquist and Estabrook  \cite{WaEs}. However, one should be aware of
an ingenious observation  that B\"acklund transformations can be reinterpreted as increments in additional discrete variables \cite{LeBe} and the difference-differential or fully discrete equations obtained in this way are integrable.
With this observation
 the nonlinear superposition principle for the potential KdV  $\x _{t}=6 (\x _{y})^2-\x _{yyy}$  presented in \cite{WaEs} can be reinterpreted as a difference equation, namely the lpKdV \mref{xi}.
%\begin{equation}
%\label{lpKdV}
%\begin{array}{l}
%\x (m+1,n+1)=\x(m,n)+{\dis \frac{k_{1}^2-k_{2}^2}{\x (m+1,n)-\x (m,n+1)}}
%\end{array}
%\end{equation}
%which is  referred to as lattice potential KdV \cite{NQC,NC}.

\begin{figure}[h]
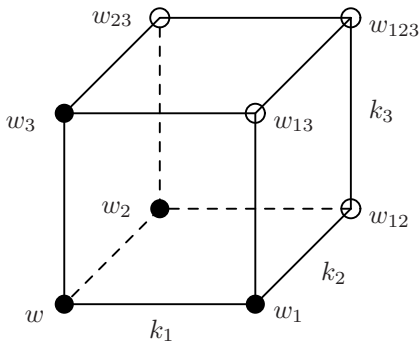

\centertexdraw{ \setunitscale 1. \linewd 0.01 \move (0 0) \linewd
0.01 \lpatt(0.05 0.05)
\lvec(0.0 1.0)
\lpatt()
\lvec (1.0 1.0)
\linewd 0.01
 \lvec (1.0 0.0)
\lpatt(0.05 0.05)
 \lvec(0 0)
\lvec(-0.5 -0.5) \lpatt()  \lvec(-0.5 0.5)  \linewd 0.01
 \lvec(0 1.0) \move (1. 0)  \linewd 0.01 \lvec(0.5 -0.5) \lvec(-0.5 -0.5)
\move(1.0 1.0) \lvec(0.5 0.5) \lvec(-0.5 0.5) \move(0.5 0.5)
\lvec(0.5 -0.5)

\move(0 0) \fcir f:0.0 r:0.05 \move(-0.5 0.5) \fcir f:0.0 r:0.05
\move(0.5 -0.5) \fcir f:0.0 r:0.05 \move(-0.5 -0.5) \fcir f:0.0
r:0.05 \move(1 1) \lcir r:0.05 \move(0.5 0.5) \lcir r:0.05 \move(0
1) \lcir r:0.05 \move(1 0) \lcir r:0.05

 \htext (-0.35 0.95)
{$\x_{23}$} \htext (-0.3 -0.05) {$\x_{2}$} \htext (1.1 0.9)
{$\x_{123}$} \htext (-0.8 0.4) {$\x_{3}$}
\htext (0.6 -0.6) {$\x_{1}$} \htext (1.1 -0.1) {$\x_{12}$} \htext
(-0.7 -0.6) {$\x$} \htext (0.6 0.4) {$\x_{13}$} \htext (0.85 -0.4)
{$k_2$} \htext (-0.05 -0.7) {$k_1$} \htext (1.1 0.45)
{$k_3$}

} \caption{3-dimensional consistency}
\label{fig:3D}
\end{figure}

Going further in reinterpretation of the paper by  Wahlquist and Estabrook
one can consider $\mathbb{Z}^3$ lattice with three copies of equation \mref{xi} given on it
\begin{equation}
\label{lpKdV3}
\begin{array}{l}
\x (m+1,n+1,s)=\x (m,n,s)+{\dis \frac{p-q}{\x (m+1,n,s)-\x(m,n+1,s)}} \\ [3mm]
\x (m+1,n,s+1)=\x(m,n,s)+{\dis \frac{p-\lambda}{\x (m+1,n,s)-\x(m,n,s+1)}} \\ [3mm]
\x (m,n+1,s+1)=\x(m,n,s)+{\dis \frac{q-\lambda}{\x (m,n+1,s)-\x(m,n,s+1)}}
\end{array}
\end{equation}
There are three different ways in which we can obtain $\x_{123}$ (according to our notation \\ $\x_{123} \equiv \x (m+1,n+1,s+1)$) in terms of $\x $, $\x_1$,
$\x _2$, $\x _3$.
Due to the suitable choice of form of the numerators in formulas \mref{lpKdV3}, no-matter which path we follow,  the resulting value $\x _{123}$ is the same  (see figure \ref{fig:3D}).
This property is referred  to as 3D-consistency or consistency around the cube \cite{ABS,FN,FN-KN}. We have \cite{WaEs}:
\begin{equation}
\label{tetr}
\x_{123}=\frac{p\x_{1}(\x_{2}-\x_{3})+q\x_{2}(\x_{3}-\x_{1})+
\lambda \x_{3}(\x_{1}-\x_{2})}{\x_{1}(\x_{2}-\x_{3})+\x_{2}(\x_{3}-\x_{1})+\x_{3}(\x_{1}-\x_{2})}.
\end{equation}
At this moment
the  independence of $\x_{123}$ on $\x $ in formula \mref{tetr} should be underlined. This property is referred  as the tetrahedron property since equation \mref{tetr} relates four points of the 3D-lattice only.
%As we shall {\color{red} see equation \mref{realcub} and models G1 and F1} does not possess tetrahedron property.

% Now we arrive at the point where advantage of difference equations over the differential ones is visible.
Finally, one can interpret the  two last equations of \mref{lpKdV3}  as a  B\"acklund transformation  for the first one.
It  is necessary to mention that both the last equations of \mref{lpKdV3} are first order fractional linear difference equations (discrete Riccati equations), when regarded as recurrence relations in the variable $s$, hence their solving can be reduced to solving linear equations only.

Extensions of the ABS list, interesting structures closely related to the consistency-around-the-cube property and solutions of the equations from the list, have been studied since the publication of paper \cite{ABS} (see
\cite{CField-general,QNCL,FN-KN,hyd-general,TsaWo-general,C.M.Viallet-general,SLN-Lagrangian,ABS-Lagrangian,Kon-Sol,Jo-Sol,N.At-Sol,Zh-Sol,NAH-Sol,AN-Sol,HZ-Sol,AHNQ3-Sol,James,RG1-Sol,RG2-Sol}).

Recapitulating, for the system \mref{lpKdV3} we have
\begin{enumerate}
\item Each of the equations include two parameters
\item Due to specific dependence on the parameters the system is compatible (consistent around the cube)
\item The two last equations of \mref{lpKdV3} are fractional linear recurrence relations in the variable $s$
\item Formula  \mref{tetr} relates $\x _{123}$ with $\x _{1}$ $\x_{2} $ and $\x _{3}$ only, the so called tetrahedron property.
\end{enumerate}
Equation \mref{realcub} and models G1 and F1   possess  properties 1. and 2. only.
It means for instance that equation \mref{realcub}
is not in the form assumed in
\cite{ABS,ABS2}, namely it is not in the form $Q(x_{12},x_1,x_2,x)=0$ where $Q$ is a multiaffine polynomial i.e. polynomial of degree one in each argument.

%%%%%%%%%%%%%%%%%%%%%%%%%
\section{Involutive maps}
%%%%%%%%%%%%%%%%%%%%%%%%%
\label{yb}

There are several procedures  for obtaining integrable mappings  \cite{ves2} from integrable lattice equations \cite{pap1,nij55,vk:rq,Spicy,BoSu-book,ABSf,pap2}.
Following the procedure of \cite{ABSf,pap2}  one can get involutive maps  which are set-theoretical solutions of the Yang-Baxter equation the so-called
Yang-Baxter maps \cite{drin,Ves,ABSf,PSTV}.
 We complement this procedure with a systematic way of finding lattices
from involutive mappings.

\begin{figure}[h]
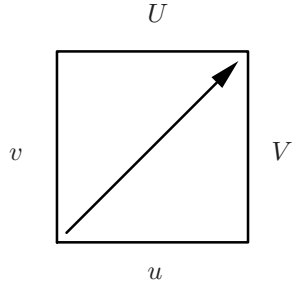

\centertexdraw{\setunitscale 0.5
\linewd 0.03 \arrowheadtype t:F
\move (-1 -1)
\lvec (-1 1) \lvec (1 1) \lvec (1. -1) \lvec(-1 -1)
\htext (-0.05 -1.4) {$u$}
\htext (-1.5 -0.15) {$v$}

\htext (-0.05 1.3) {$U$}
\htext (1.25 -0.15) {$V$}
\move(-0.9 -0.9)  \avec (0.9 0.9)
}
\caption{A map on $\mathbb{C}^2$.}
\label{fig:c2-c2}
\end{figure}

We confine ourselves to a family of involutive maps $\mathbb{C}^2 \ni (u,v) \mapsto (U,V) \in \mathbb{C}^2 $ (see figure \ref{fig:c2-c2}) which are related to the Yang-Baxter map denoted by
$F_V$ in the classification list of quadrirational Yang-Baxter maps given in \cite{ABSf}.
\begin{equation}
\label{fv}
\begin{array}{l}
(F_V): \qquad \qquad U=v+{\dis \frac{p-q}{u-v}}, \qquad  V=u+{\dis \frac{p-q}{u-v}}
\end{array}
\end{equation}
Apart from $F_V,$  we number among the family,
the Yang-Baxter map denoted by $H_V$ in article \cite{PSTV} (c.f.  \cite{Ves})
\begin{equation}
\label{hv}
\begin{array}{l}
(H_V): \qquad \qquad U=v+{\dis \frac{p-q}{u+v}}, \qquad  V=u-{\dis \frac{p-q}{u+v}}
\end{array}
\end{equation}
and
its companion (i.e. the map $(u,v)\to(U,V)$ that arises from the switching of $v$ and $V$ in \mref{hv})
\begin{equation}
\label{chv}
\begin{array}{l}
(cH_V): \qquad \qquad U=-v+{\dis \frac{p-q}{u-v}}, \qquad V=-u+{\dis \frac{p-q}{u-v}}.
\end{array}
\end{equation}
Note that (\ref{chv}) is not a Yang-Baxter map \cite{PSTV}, and in order to illustrate that the Yang-Baxter property is not crucial from the point of view of the procedures described in this paper,  we deal here with the map \mref{chv} mainly.

\begin{figure}[h]
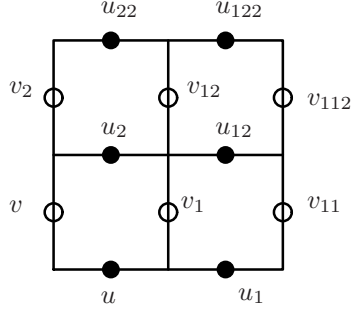

\centertexdraw{\setunitscale 0.6
\linewd 0.02
\move (-1 -1)
 \lvec (-1 1)  \lvec (1 1) \lvec (1 -1) \lvec (-1 -1)
 \move (0 -1)
 \lvec (0 1)
 \move (-1 0)
 \lvec (1 0)
\move(0 -0.5) \lcir  r:0.08
\htext (0.1 -0.5) {$v_1$}

\move(-0.5 -1) \fcir f:0  r:0.08
\htext (-0.6 -1.3) {$u$}

\move(0.5 -1) \fcir f:0  r:0.08
\htext (0.6 -1.3) {$u_1$}

\move(-0.5 1) \fcir f:0  r:0.08
\htext (-0.6 1.2) {$u_{22}$}

\move(0.5 1) \fcir f:0  r:0.08
\htext (0.4 1.2) {$u_{122}$}

\move(-1 0.5) \lcir  r:0.08
\htext (-1.4 0.5) {$v_2$}

\move(-1 -0.5) \lcir  r:0.08
\htext (-1.4 -0.5) {$v$}

\move(1 0.5) \lcir  r:0.08
\htext (1.2 0.4) {$v_{112}$}

\move(1 -0.5) \lcir  r:0.08
\htext (1.2 -0.5) {$v_{11}$}

\move(-0.5 0) \fcir f:0  r:0.08
\htext (-0.6 0.15) {$u_2$}

\move(0.5 0) \fcir f:0  r:0.08
\htext (0.4 0.15) {$u_{12}$}

\move(0 0.5) \lcir  r:0.08
\htext (0.15 0.5) {$v_{12}$}

}
\caption{The map as a lattice.}
\label{fig:building-latt}
\end{figure}

The standard procedure for reinterpretation of a map as equations on a lattice is based on the identification (see Figure \ref{fig:building-latt}.)
\begin{equation}
\label{id}
u=u(m+{ \frac{1}{2}},n),\quad v=v(m,n+{ \frac{1}{2}}), \quad U=u(m+{ \frac{1}{2}},n+1), \quad V=v(m+1,n+{ \frac{1}{2}}).
\end{equation}
%where the function $u(m+{ \frac{1}{2}},n)$ is given on horizontal edges only and $v(m,n+{ \frac{1}{2}})$ is given on vertical ones.
For the map \mref{chv} we get exactly the system (\ref{uv}).
%\begin{equation}
%\label{chvl}
%u(m,n+1)
%=-v(m,n)+{\dis \frac{p(m)-q(n)}{u(m,n)-v(m,n)}} \quad v(m+1,n)=-u(m,n)+{\dis \frac{p(m)-q(n)}{u(m,n)-v(m,n)}}.
%\end{equation}

It is  important for us to find functions $F$ and $G$ such that
\begin{equation}
\label{FGfg}
F(U)+G(V)= f(u)+g(v).
\end{equation}
In the case   of the map \mref{chv}
it leads to an equation in $F$ and $G$ (we differentiate \mref{FGfg} with respect to $u$ and~$v$)
\begin{equation}
\label{long}
F''(U)[(U-V)^2-(p-q)]-2 F'(U)(U-V)-G''(V)[(U-V)^2+(p-q)]-2 G'(V)(U-V)=0.
\end{equation}
 To find a solution  of equation \mref{long} that is valid for all values of $U$ and $V$ we observe that the  necessary conditions that the functions $F$ and $G$ should satisfy are
\begin{equation}
\begin{array}{l}
F''(U) [U^2-(p-q)]-2 U F'(U)=c_1U^2+c_2U +c_3\\
G''(V) [V^2+(p-q)]-2 V G'(V)=c_4V^2+c_5V +c_6
\end{array}
\end{equation}
where the $c_i$ are some constants. Solving the ODEs for F(U) and G(V) and plugging the results into equation \mref{long}
we find the  following general solution of \mref{long}
\begin{equation}F(U)+G(V)=c_1 (U-V)+c_2 (U^2-V^2)+c_3 (U^3+pU-V^3-qV)+c_4.\end{equation}
As a result we arrive at the following integrals with additively separated variables
\begin{equation}
\label{Hh1}
{\mathcal H}_1 (u,v)=u-v, \quad  {\mathcal H}_3 (u,v)=u^3+p u -v^3-q v,
\end{equation}
and a 2-integral with separated variables as well
\begin{equation}
\label{H2}
{\mathcal H}_2 (u,v)=u^2 - v^2.
\end{equation}
After the reinterpretation \mref{id}  of  the map as a system of equations on a lattice,  integrals \mref{Hh1} and \mref{H2}
will give rise to equations that  guarantee the  existence of potentials.

The question arises whether one can
extend the system \mref{uv} to a  compatible multidimensional  system on the ${\mathbb Z}^n$ lattice. The answer is positive and we will take up this issue in the next section.

We end the section with a proposition
of an equivalence relation in the set of 2D maps. According to the proposition all three maps \mref{fv}, \mref{hv} and \mref{chv} are  equivalent.

\begin{prop}
\label{p1}
%{\em \bf Proposition 1}
Two 2D maps are equivalent if their  systems arising from identification \mref{id}, let say on function $\tilde{u}^i$ and ${u^i}$, can be related  by an invertible point transformation:
\begin{displaymath}
\begin{array}{l}
\tilde{u}^i=F^i(u^1,u^2,m,n), \quad i=1,2, \qquad
\tilde{m}=m, \qquad
\tilde{n}=n.
\end{array}
\end{displaymath}
\end{prop}

%%%%%%%%%%%%%%%%%%%%%%%%%%%%%%%%%%%%%
\section{Idea  $I_V$ and its idolons}
%%%%%%%%%%%%%%%%%%%%%%%%%%%%%%%%%%%%%
\label{main}
We consider the $\mathbb{Z}^n = \left\{(m_1,\ldots,m_n) \, :\, m_i \in \mathbb{Z}, i=1,\ldots, n \right\}$ lattice together with its edges i.e. segments connecting two consecutive points. The edges can be viewed as (and described by) a pair of vertices. We refer to the  elements of the set of ordered pairs of points \newline
$\left\{ ((m_1,\ldots, m_i,\ldots ,m_n),(m_1,\ldots, m_i+1,\ldots ,m_n)) \, :\, m_i \in \mathbb{Z}, i=1,\ldots, n  \right\}$, as  edges in the i-th  direction.  We denote by subscript the forward shift in the i-th direction $T_i$  i.e. \\
$T_if(m_1,\ldots, m_i,\ldots ,m_n)\equiv f_i(m_1,\ldots, m_i,\ldots ,m_n)=f(m_1,\ldots, m_i+1,\ldots ,m_n)$.
We omit independent variables to make the formulas shorter (e.g. $T_i f \equiv f_i$).
By $\Delta_i$ we
denote the forward difference operator $\Delta_i=T_i -1$.

We take into consideration $n$ functions $u^i$,  $i=1,\ldots, n$ (mind that we enumerate functions by a superscript!). The i-th function $u^i$ is given on edges in the i-th direction only.
A function ${\mathbf u}= (u^1,\ldots,u^n)$ from the set of edges to ${\mathbb R}^n$ (or ${\mathbb C}^n$) can be viewed as a vector field on a lattice.
We are searching for  functions that obey the following system of difference equations
\begin{equation}
\label{ui}
(I_V): \qquad \qquad \qquad \qquad u^i_j+u^j=\frac{p^i-p^j}{u^i-u^j} \qquad i,j=1,\ldots, n \qquad i \neq j
\end{equation}
where the $p^i$ are given functions of the i-th argument only ($p^i(m_i+\frac{1}{2})$).

To emphasise the meaning of the system  \mref{uv} and its multidimensional extension \mref{ui}
and the fact that the system can have various potential ``reflections''
we suggest to adopt Plato's terminology and refer to the system as the {\em Idea  system }
(or the representative of an  {\em Idea}, see Proposition \ref{p2}) associated with the map \mref{chv}  and to underline the connection with \mref{chv} we denote it by $I_V$.
Whereas the models
H1, G1 and F1 (and their higher dimensional extensions, see below) we refer to as idolons of the Idea system
(keeping the original ancient Greek form of the word idol and having in mind rather idols  from Bacon's works than any contemporary meaning of the word).
Note that in this terminology we do not exclude the case where a given idolon  participates in
several Ideas.

The following facts hold
\begin{itemize}
\item  system (\ref{ui}) is compatible i.e.
\begin{equation}
\label{com}
u^i_{jk}=u^i_{kj} \qquad i,j,k =1,\ldots, n \qquad i \neq j \neq k \neq i
\end{equation}
Prescribing the value of  each function $u^i$ on the $i$-th initial condition line ($m_j=0$ for $j\neq i$) and using \mref{ui} one can find recursively $u^i$ in the whole domain (singularities can occur because of the vanishing of the denominator on the right hand side of \mref{ui}).
\item The following equality holds
\begin{equation}
\label{ux}
\Delta_j u^i=\Delta_i u^j \qquad i,j =1,\ldots, n \qquad i \neq j
\end{equation}
which implies the existence of a potential $x$ (given on vertices of the lattice) such that
\begin{equation}
\label{x}
 u^i= \Delta_i x \qquad i =1,\ldots, n
\end{equation}
In terms of the potential $x$ the system \mref{ui} reads
\begin{equation}
\label{xe}
 (x_{ij}-x)(x_i-x_j)=p^i-p^j \qquad i,j =1,\ldots, n \qquad i \neq j
\end{equation}
so we  get  the system  of lpKdV equations.
Since the existence of $x$  is guaranteed for any initial data for $x$ (excluding the ones that leads to $u^i-u^j=0$) on initial lines, system \mref{xe} is nD-consistent.
\item From  system \mref{ui} one infers that
\begin{equation}
\label{zf}
(T_j+1) \left[(u^i)^2-p^i \right] =(T_i+1) \left[(u^j)^2-p^j \right] \qquad i,j =1,\ldots, n \qquad i \neq j
\end{equation}
Equations \mref{zf} imply the existence of a potential $z$ (given on vertices of the lattice) such that
\begin{equation}
\label{z}
 (u^i)^2-p^i =  (T_i+1) z \qquad i =1,\ldots, n
\end{equation}
In terms of potential $z$ the system \mref{ui} can be rewritten as
\begin{equation}
\label{ze}
z_{ij}-z= (p^i-p^j) \frac{z_j-z_i}{(u^i-u^j)^2}  \qquad i,j =1,\ldots, n \qquad i \neq j.
\end{equation}
We get the system of equations \mref{ui}, \mref{z} and  \mref{ze}, which we refer to as system of G1 models and which is nD-consistent.
An explicit manifestation of 3D-consistency is the formula
\begin{displaymath}
\begin{array}{l}
z_{ijk}+z=\\
\frac{[p^i u^i  (u^j-u^k) + p^j  u^j(u^k-u^i)+p^k u^k (u^i- u^j)]^2}{[p^i(u^j-u^k)+p^j(u^k-u^i)+p^k(u^i-u^j)]^2}+
\frac{[u^i(p^j-p^k)(p^j+p^k-p^i)+u^j(p^k-p^i)(p^i+p^k-p^j) +u^k(p^i-p^j)(p^i+p^j-p^k)]}{p^i(u^j-u^k)+p^j(u^k-u^i)+p^k(u^i-u^j)}.
\end{array}
\end{displaymath}
Finally, we infer that $z$ must satisfy
\begin{equation}
%\label{}
\begin{array}{l}
(z_{ij}-z)^2 (z_i-z_j+p^i-p^j)^2 +(p^i-p^j)^2 (z_i-z_j)^2+\\
 2(z_{ij}-z)(p^i-p^j)(z_i-z_j)(z_i+z_j+2 z + p^i + p^j)=0
\end{array}
\end{equation}
compare with \cite{ves3}.
\item The third identity giving rise to a potential is
\begin{equation}
\label{uf}
\Delta_j \left[(u^i)^3+a^i u^i\right] =\Delta_i \left[(u^j)^3+a^j u^j \right] \qquad i,j =1,\ldots, n \qquad i \neq j
\end{equation}
where $a^i$ are functions of  the i-th argument only ($a^i(m_i+\frac{1}{2})$) such that
\begin{equation}
\label{a}
a^i-a^j=3(p^j-p^i) \qquad i,j =1,\ldots, n \qquad i \neq j
\end{equation}
In fact by the introduction of the $a^i$ functions we tacitly smuggled in the possibility of combination of integrals \mref{Hh1}.
Equations \mref{uf} imply the existence of a potential $f$ (given on vertices of the lattice again) such that
\begin{equation}
\label{f}
 (u^i)^3+a^i u^i= \Delta_i f \qquad i =1,\ldots, n
\end{equation}
In terms of the potential $f$ the system \mref{ui} can be rewritten as
\begin{equation}
\label{fe}
f_{ij}-f= (p^i-p^j)\left[\frac{(p^i-p^j)^2}{(u^i-u^j)^3}+\frac{f_i-f_j}{(u^i-u^j)^2} -u^i+u^j\right] \qquad i,j =1,\ldots, n \qquad i \neq j.
\end{equation}
Just like in the previous cases we get the system of equations \mref{ui}, \mref{f} and \mref{fe}, which we refer to as system of F1 models, system which is nD-consistent.
The 3D-consistency is manifested in the formula

{\small
\begin{displaymath}
\begin{array}{l}
f_{ijk}-f=
\frac{(p^i u^i  u^{jk} + p^j  u^j u^{ki}+p^k u^k u^{ij})^3}{(p^i u^{jk}+p^j u^{ki}+p^k u^{ij})^3}
-
\frac{3 p^{ij} p^{jk} p^{ki} }{p^i u^{jk}+p^j u^{ki}+p^k u^{ij}}+\\
\frac{
3\left\{
p^i u^i  p^{ij} p^{ki} [(u^j)^2 + (u^k)^2] +
p^j u^j  p^{ij} p^{jk} [(u^k)^2 + (u^i)^2] +
p^k u^k  p^{jk} p^{ki}  [(u^i)^2 + (u^j)^2] +
2 u^i u^j u^k [(p^{ij}-p^k)(p^{jk}-p^i)(p^{ki}-p^j)+p^ip^jp^k]
\right\}}{[p^i u^{jk}+p^j u^{ki} +p^k u^{ij}]^2}
\end{array}
\end{displaymath}
}
where for the sake of brevity we have introduced  notation $u^{ij}:=u^i-u^j$ and $p^{ij}:=p^i-p^j$.
The real case with positive parameters $a^i$ provides us with an example of a system of non-multiaffine consistent-around-the-cube lattice equations \mref{realcub}.

\item The proper initial value problem for  systems of models H1, G1 and F1 is to prescribe a value of a potential at the intersection of initial lines and  values of $u^i$ on the edges that belong to the initial lines. Then we have a unique solution to the models. However, we find it interesting to discuss
the case when the values of potentials are given on the initial lines.
 In the case of complex-valued functions,   we can find from equations \mref{x}, \mref{z} or \mref{f},
respectively at most $k=1,2$ or $3$  values of $u^i$
on an edge belonging to the initial line. Therefore  for an elementary n-cube  having prescribed values of potential at a given vertex and its closest neighbours 
one can  get at most $k^n$ values of the potential at the farthest (from the given one) vertex.

\item
One can combine all of the cases above by introducing the family of potentials $\psi$
\begin{equation}
c_0+c_1u^i  +c_2 (-1)^{m_1+\ldots +m_n}[(u^i)^2-p^i]+ c_3 [(u^i)^3+p^i u^i] =\psi_i-\psi \qquad i =1,\ldots, n
\end{equation}
We postpone the investigation  of the lattice equation on $\psi$, which includes four parameters $c_0$, $c_1$, $c_2$ and $c_3$, to a forthcoming paper \cite{PKMN3}.

\item Eliminating $u^i$ from \mref{z} and from \mref{f} by means of \mref{x} we get
\begin{equation}
\label{xz}
(T_i+1) z= ( \Delta_i x)^2- p^i  \qquad i =1,\ldots, n
\end{equation}
\begin{equation}
\label{xf}
\Delta_i f= (\Delta_i x)^3+a^i \Delta_i x  \qquad i =1,\ldots, n
\end{equation}
from which we infer
\begin{itemize}
\item If $x$ obeys $\Delta_i \Delta_j x=0 $ then $f$ given by \mref{xf} obeys $\Delta_i \Delta_j f=0 $ as well.
\item If $x$ obeys \mref{xe} then $f$ given by \mref{xf} obeys \mref{fe}.
\item If $x$ obeys \mref{xe} then $z$ given by \mref{zf} obeys \mref{ze}.
\end{itemize}
\item If we replace $(p^i,u^i)$ with $(-p^i, (-1)^{m_1+\ldots +m_n}u^i)$ we get
\begin{equation}
\label{if}
u^i_j-u^j=\frac{p^i-p^j}{u^i-u^j}
\end{equation}
In the two-dimensional case, after making the identification \mref{id}, one obtains the $F_V$ map \mref{fv}. The whole procedure we have described so far can be repeated using the $F_V$ map instead of the $cH_V$ map.
Since ``entities must not be multiplied beyond necessity''  we arrive at the proposition.

\begin{prop}
\label{p2}
 Two idea systems, let's say on functions $\tilde{u}^i$ and ${u^i}$, are equivalent iff they are related by an invertible point transformation
\begin{equation}
\label{point}
\tilde{u}^i=F^i(u^1,\ldots,u^n,m_1,\ldots,m_n), \quad \tilde{m}_i=m_i, \quad i=1,\ldots,n\\
\end{equation}
 An equivalence class of idea systems we refer to as Idea.
\end{prop}

Idea systems \mref{ui} and \mref{if} are representatives  of the same Idea, the Idea associated with the family of maps discussed in previous section. We denote the Idea by $I_V$ just like its representative \mref{ui}.

\item
If we replace $u^2$ with $-u^2$ we get equations
\[u^i_2=u^2+\frac{p^i-p^2}{u^i+u^2}, \qquad  u^2_i=u^i-\frac{p^i-p^2}{u^i+u^2} \quad i \neq 2 \]
which in the two-dimensional case, after making the identification \mref{id}, changes to the $H_V$ map \mref{hv}. However, it is not possible to turn all equations  into the form
 $u^j_i=u^i-\frac{p^i-p^j}{u^i+u^j}$. In this case $u^j_{ik}\neq u^j_{ki}$.
\end{itemize}

%%%%%%%%%%%%%%%%%%%%%%
\section{Comments}
%%%%%%%%%%%%%%%%%%%%%%
\label{comm}
We focused in this paper on two models G1 and F1 which are multidimensionally consistent.  The models are related to the system of lattice potential KdV equations \mref{xe} by non-auto-B\"acklund transformations \mref{xz} and \mref{xf}. Equation \mref{realcub} and its multidimensional extension is a particular (real) case of F1. Because of the lack of the tetrahedron property, for $n>2$,  equation \mref{realcub} cannot be related by a point transformation to any
system of equations from Adler-Bobenko-Suris list \cite{ABS}.
An open question is whether one can find difference substitution (i.e. a generalization of point transformations \mref{point} where functions $F^i$ can depend on functions $u^i$ given in several points) that reduce solving system \mref{ze} or system \mref{fe} to solving equations from the Adler-Bobenko-Suris list  or to solving linear equations. We shall pick up this issue in the near future.

There is no doubt that the notion of consistency (compatibility) is an important ingredient of the theory of integrable systems. However understanding consistency itself as integrability can lead to confusion.
One of the main features of integrable systems  is that their solvability can be reduced to solving linear equations.
Leaving the assumption of a multiaffine form of the quad-graph equation we face {\em terra incognita}.
One cannot exclude the situation when the consistent equation cannot be linked to any integrable equation by a linearizable transformation.
It can produce equations with essentially nonlinear B\"acklund transformations
and that is why the conception of consistency around the cube is interesting in itself. We take the stand that we should not mix it with integrability at this stage. The class of integrable systems may be  identical with the class of consistent-around-the-cube equations or may be not.

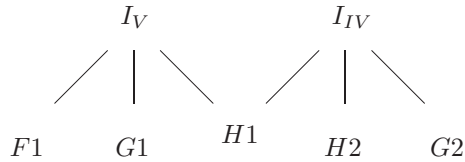
\begin{figure}[!ht]
\begin{picture}(60,60)(-140,10)
\put(45,60){$I_V$}  \put(125,60){$I_{IV}$}
\put(40, 50){\line(-1, -1){20}} \put(50, 50){\line(0, -1){20}} \put(60, 50){\line(1, -1){20}}
\put(120, 50){\line(-1, -1){20}}
\put(130, 50){\line(0, -1){20}}
\put(140, 50){\line(1, -1){20}}
\put(3,10){$F1$}  \put(43,10){$G1$}  \put(83,15){$H1$}
\put(122,10){$H2$} \put(162,10){$G2$}
\end{picture}
\caption{Ideas $I_{IV}$ and $I_V$ and their idolons}
\label{pic}
\end{figure}

For now we are going to confine ourselves to systems that are linked to the ABS  list by a linearizable B\"acklund or Miura transformation (see e.g. \cite{James,Decio}) i.e. by a fractional linear difference equation. So we are going to extend  table 3
of paper \cite{James} (see also \cite{Decio})  to non-multiaffine cases. In a forthcoming paper we deal with  the simplest case when the link is given by discrete
quadratures (so to say ``simplest'' fractional linear transformation).
It will also cover multiplicative cases, for instance the substitution
\begin{equation}
\begin{array}{l}
u=x_1 x, \quad v=x_2 x
\end{array}
\end{equation}
and after making the identification \mref{id} changes the lpKdV equation \mref{xi} into the Yang-Baxter map $F_{IV}$  from the list given in  \cite{ABSf}
\begin{equation}
V=u\left(1+\frac{p-q}{u-v} \right), \quad
U=v\left(1+\frac{p-q}{u-v} \right).
\end{equation}
The most general function with property \mref{FGfg} for this map is a linear combination of functions
{\small
\begin{equation}
\begin{array}{l}
\!\!\!
\!\!\!\!\!\!
\log \frac{U}{V}= \log \frac{v}{u}, \quad
U-V=-(u-v+p-q), \quad
U^2-V^2+2pU-2qV=-(u^2-v^2+2pu-2qv +p^2-q^2)
\end{array}
\end{equation}}
and a constant function.
It leads to the following potentials $x$, $y$ and $z$
\begin{equation}
\begin{array}{l}
u=x_1 x, \quad v=x_2 x\\
2u+p=y_1+y, \quad 2v+q=y_2+y\\
u^2+2pu+\frac{1}{2}p^2=z_1+z, \quad v^2+2qv+\frac{1}{2}q^2=z_2+z
\end{array}
\end{equation}
The first potential  satisfies the lpKdV equation  \mref{xi}, the second potential satisfies equation $H2$ from the ABS list \cite{ABS}
\hbox{$(y_{12}-y) (y_{1}-y_{2}) -(p-q)(y_{12}+y_{1}+y_{2}+y)+p^2-q^2=0$}
and the third one gives rise another non-multiaffine model we denote tentatively denote by $G2$ for equivalence and classification in the class of non-multiaffine equations is yet to be defined
\begin{equation}
\begin{array}{l}
 z_{12}-z=
\frac{p-q}{u-v} \left[2uv+\frac{1}{2} (p+q)(u+v)+uv\frac{p-q}{u-v}\right].
\end{array}
\end{equation}
The Idea associated with $F_{IV}$ is (we denote it by $I_{IV}$)
\begin{equation}
\begin{array}{l}
(I_{IV}):\qquad  \quad \quad   u^i_j=u^j \left(1+\frac{p^i-p^j}{u^i-u^j}\right) \qquad i,j=1,\ldots, n \qquad i \neq j
\end{array}
\end{equation}
We see that idolon $H1$ participates both in the $I_{IV}$ and $I_V$ Ideas which we illustrate in the final figure, Figure~\ref{pic}.

{\bf Acknowledgements}
We are grateful to the organizers of the
Isaac Newton Institute for Mathematical Sciences, programme of
Discrete Integrable Systems, for giving us the opportunity to take part in this event, extraordinary in every way, shape and form.
We are also grateful for the support of the organizers of the 9th SIDE meeting in Varna, where the paper gained its tentative form.
MN would like to express  gratitude to Tomasz Steifer for fruitful discussions on Plato's ideas and idolons and referring us to the Liddell \& Scott Greek lexicon  and Bacon's works. MN also thanks Paolo Maria Santini, Decio Levi and Frank Nijhoff not only for literature guidelines but most of all for encouragement when our ardour began to cool.  P.K. thanks Tasos Tongas for his useful comments and for referring us to  article \cite{ves3}. Finally, our special thanks to Sarah Lobb for Englishization of this paper.

\bibliographystyle{unsrt}	
\bibliography{PKMN-paper1-ref}

\end{document}